\documentclass[jphysg,11pt]{article}
\usepackage[dvips]{graphics}
\usepackage{epsfig}

\def\lapproxeq{\lower .7ex\hbox{$\;\stackrel{\textstyle<}{\sim}\;$}}
\def\gapproxeq{\lower .7ex\hbox{$\;\stackrel{\textstyle>}{\sim}\;$}}

\title{Fine Structure in the Gamma Ray Sky}

\author{A. D. Erlykin$^{1,2}$ and A. W. Wolfendale$^2$\\[1ex] $^1$PN Lebedev
Physical Institute, Moscow, Russia\\ $^2$Department of Physics,
University of Durham, Durham, UK}

\date{~}

\begin{document}

\def\gtrsim{ \;\raisebox{-.7ex}{$\stackrel{\textstyle
>}{\sim}$}\; }
\def\lesim{ \;\raisebox{-.7ex}{$\stackrel{\textstyle
<}{\sim}$}\; }

\maketitle

\begin{abstract}
The EGRET results for gamma ray intensities in and near the
Galactic Plane have been
analysed in some detail.  Attention has been concentrated on
energies above 1 GeV and the individual intensities
in a $4^{\circ}$ longitude bin have been determined and compared 
with the large scale mean found from a nine-degree polynomial fit.

Comparison has been made of the observed standard deviation for 
the ratio of these intensities with
that expected from variants of our model.  The basic model adopts
cosmic ray origin from supernova remnants, the particles then
diffusing through the Galaxy with our usual `anomalous diffusion'.
  The variants involve the clustering of SN, a frequency 
distribution for supernova explosion energies, and 'normal',
rather than 'anomalous' diffusion.

It is found that for supernovae of unique energy, and our usual anomalous diffusion, 
clustering is necessary, particularly in the Inner Galaxy. An alternative, and 
preferred, situation is to adopt the model with a frequency distribution of supernova 
energies. The results for the Outer Galaxy are such that no clustering is required.

\end{abstract}

\section{Introduction}

Although supernova remnants (SNR) are often invoked as the source
of cosmic rays (CR), of energy below the `knee' at about 3 PeV,
\cite{axfo,EW1}, there are still a number of imponderables. Here, we concentrate on two
 aspects: \\
(i) SN energies are not unique, but have a certain frequency distribution 
 involving some 'super-supernovae' \cite{svesh}. We refer to this model as SSNR. \\
(ii) SNR in the Galaxy are not distributed at random, but rather in clusters, in
both space and time and, preferentially, in spiral arms \cite{goud}.
This is NSNR model (~Normal SNR~).

In our previous work \cite{EW3} we used the acceleration model of \cite{EW1},
which assumed that particles are trapped for up to $8.10^{4}y$
after the SN explosion and that SN occurred randomly in time in
the Galaxy (at an average rate of $10^{-2}~y^{-1}$).  The distribution in
space was drawn from the usual average radially symmetrical
distribution of SN surface density, with a peak at a Galactocentric
distance of D=4 kpc and falling rapidly with increasing D in the
vicinity of the sun (~at D=8.5~ kpc~).  Here, we go on to make
allowance for the clustering of the SN - and hence the SNR - in
various ways, and, separately, consider a frequency distribution of SN energies:
 the SSNR model.

The quantity used as an indicator of CR intensity in the Galaxy is
the gamma ray intensity \cite{cill} which has been measured as a function
of longitude, latitude and gamma ray energy.  Most of the present work
relates to gamma rays above 1 GeV which are considered to be
produced predominantly by CR nuclei (mainly protons) of median energy $\sim$ 40
GeV \cite{gral}.  In zero'th order of accuracy, the mean CR intensity
along a particular line of sight ($\ell,b$) is simply proportional
to the measured gamma ray intensity divided by the column density
of target gas.  In the actual analysis we use a more elaborate
method but the basis is the same.

The method is to study the ratio of the observed gamma ray
intensity to the large scale average and use that as an indicator
of the small scale variability of CR intensity from place to
place.  The variability can then be compared with that predicted
by models involving SN, clustered and unclustered, normal and with 
different energies, using the same
technique of comparing the intensity with the average.

\section{The Basic Data}

Figure 1 gives an example of the data (~from \cite{cill}~); it refers to
gamma rays with energy above 1 GeV and having $|b|<2^{\circ}$,
i.e., it relates to the Galactic Plane. Identified discrete sources have 
been removed and
the workers estimate that less than 10\% of the remaining flux is
due to unresolved sources.
\begin{figure}[htb!]
\begin{center}
\includegraphics[height=12cm,width=8cm,angle=-90]{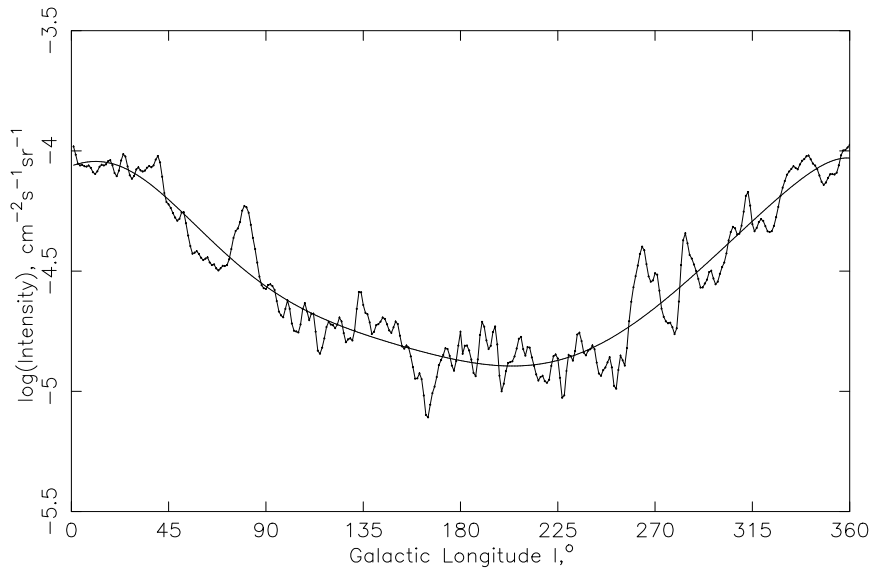}
\caption{\footnotesize  Profile of the gamma ray intensity vs
longitude, for $E_{\gamma}>1 ~GeV$ and $|b|<2^{\circ}$.  The
results are from \cite{cill}.  The standard deviations on the points due 
to Poissonian fluctuations are
typically $\pm2.3\%$ in the Inner Galaxy and $\pm4.8\%$ in the
Outer Galaxy. The experimental data are averaged over the longitude bin
$\Delta l = 1^\circ$. The smooth line is the best fit of the profile by 
the 9-degree polynomial.}
\end{center}
\label{fig:grsky1}
\end{figure}    

We have used the basic data for the numbers of gamma rays to
evaluate the statistical uncertainty in the individual
intensities.

The data have been divided into `Inner'
($\ell:270^{\circ}-350^{\circ};~10^{\circ}-90^{\circ}$) and
`Outer' ($\ell:90^{\circ}-180^{\circ}-270^{\circ}$), the innermost
region ($|\ell|<10^{\circ}$) being omitted because of confusion
there.

Two latitude ranges are considered: $|b|<2.5^{\circ}$ and $|b|:
2.5^{\circ}/5^{\circ}$. The gamma rays from the individual $b$- and $l$-ranges
 are derived from regions of the Galaxy distributed along the line of sight but
we can identify a median D-value for each range.
Restricting attention to gamma ray production in molecular clouds
with the scale height of $\sim70$ pc, the corresponding distances
are $\sim$3 kpc and 1 kpc from the sun, for $|b|<2.5^{\circ}$ and
$|b|=2.5^{\circ}/5^{\circ}$ respectively.  Allowing for production
in $HI$, with its greater scale height, the D-values will be
somewhat different (~but by not much because more of the gas -
associated fluctuations arise from the clumpy $H_{2}$ rather than
$HI$~).

\section{The Analysis}

\subsection{The Measured Fluctuations}

The `figure of merit' adopted is the ratio of the observed gamma
ray intensity to the smoothed line through the points (~9-degree
polynomial, equivalent, approximately, to a smoothing over a 20$^\circ$
longitudinal bin for the Inner and Outer Galaxy regions, separately~).  
The shape of the line is determined essentially by
the kiloparsec-scale spatial variations of the cosmic ray
intensity and the target gas. Figure 1 gives the smoothed curve
for $|b|<2^{\circ}$.

The ratios have been determined for various bins of longitude,
$\Delta\ell$, specifically, $1^{\circ},~2^{\circ},~4^{\circ}$ and
$8^{\circ}$.  Figure 2 shows the corresponding frequency
distributions of the differences between the EGRET data and the
fit. The numbers of entries, the means and the rms ($\sigma-$) values are
given in Table 1.  It is with the latter rms values that we are mainly concerned.
\begin{figure}[htb!]
\begin{center}
\includegraphics[height=7cm,width=7cm]{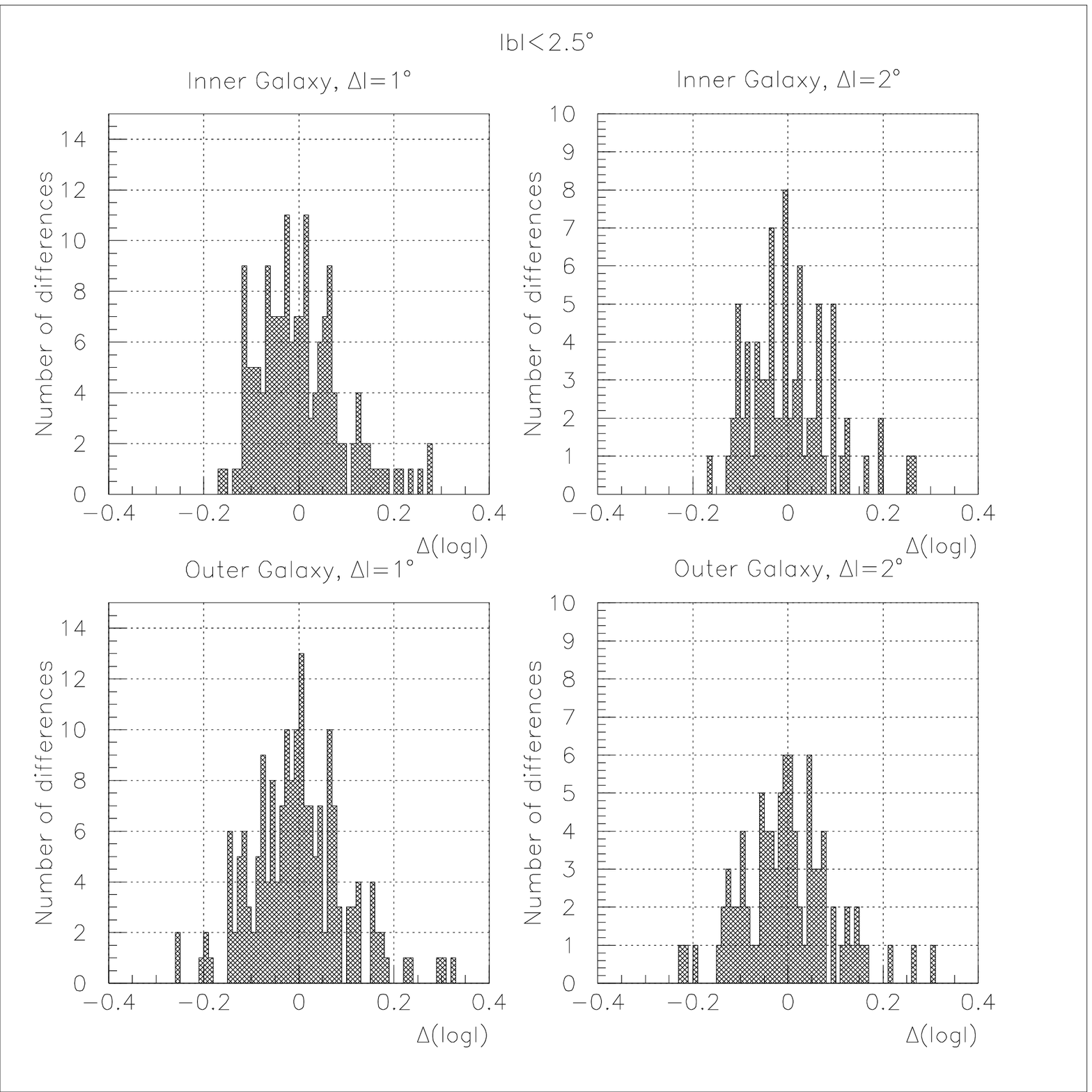}
\includegraphics[height=7cm,width=7cm]{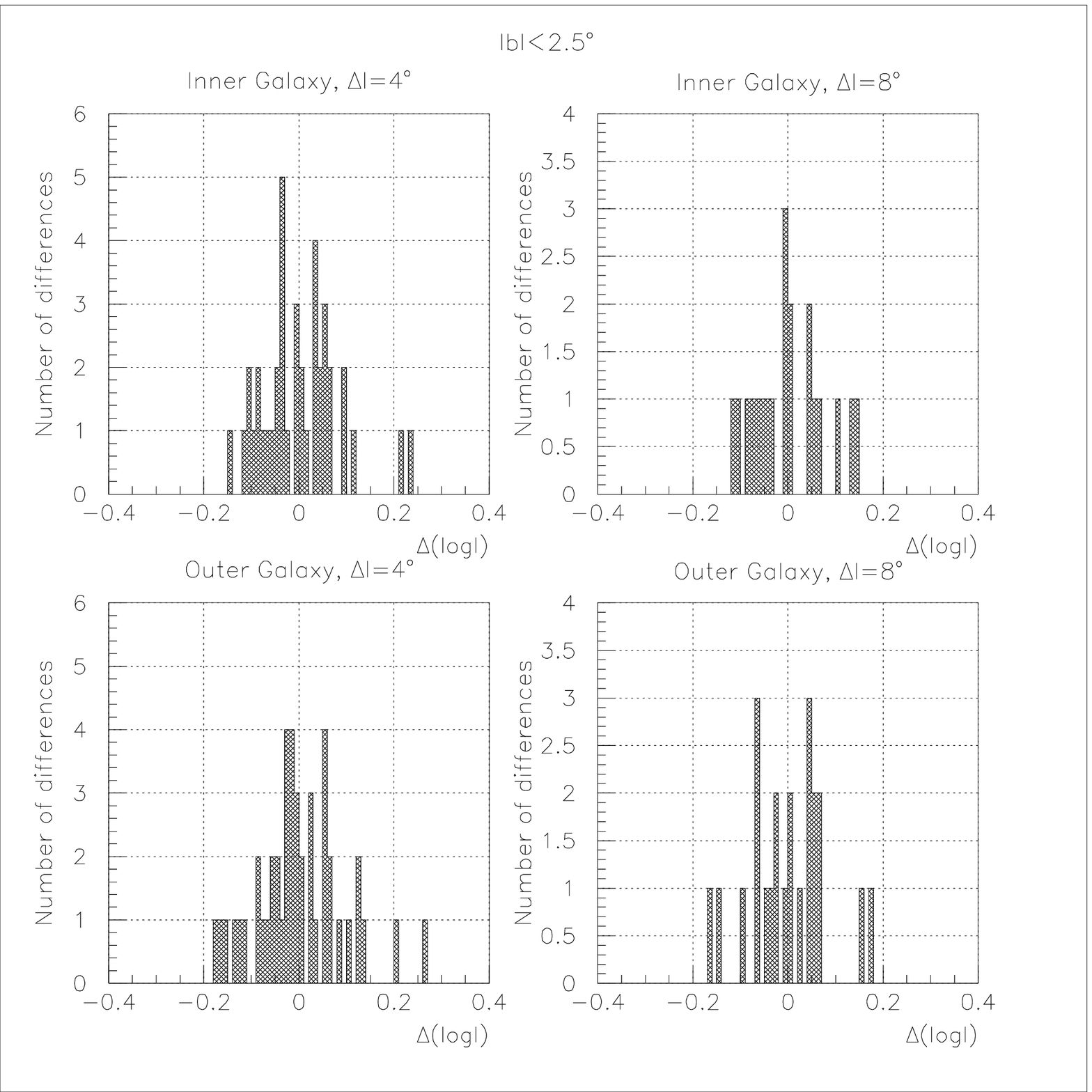}
\caption{\footnotesize The distribution of differences between the EGRET 
intensities and
the polynomial fit for various $\bigtriangleup \ell$-values. The
numbers of entries, the means and the RMS ($\sigma-$) values are
given in Table 1.}
\end{center}
\label{fig:grsky2}
\end{figure}    

The appropriate $\sigma$-value to be used is a compromise between
the $\bigtriangleup\ell$ over which intensity changes are expected
(all values but with a bias towards degree scales) and the need to
get away from small $\bigtriangleup\ell$-values, where bin-to-bin
correlations are serious, because of the finite range of latitudes
($|b|<2.5^{\circ}$) over which the averaging is made.

We consider that $\bigtriangleup\ell =4^{\circ}$ is most
appropriate and this will be used in the model calculations.

Figure 3 gives the values of $\sigma~(4^{\circ})$ for the observed
data (~after correction for noise~) and for various model calculations, 
to be described shortly.
The values are plotted at increasing (from left to right)
distances from the Galactic Centre of the median gamma ray
production regions.
\subsection{Variations in the column density of gas}
In all our models the radial distribution of gas has axial symmetry. It means that 
for a smooth non-fluctuating CR distribution the longitudinal distribution of the 
gamma-ray intensity has also to be smooth. However, we know that the ISM is highly 
non-uniform and this non-uniformity will contribute to the fine structure of
 the gamma-ray intensity profile.  

A number of facts are relevant, as follows: \\
(i) The large scale dependence of {\em N(H)} on {\em l} is automatically included 
through the radial distribution. \\
(ii) The 9-degree polynomial fit causes changes on scales of about 
$\delta l = 20^\circ$ (~and above~) to be allowed for. \\
(iii) Much of the clumpiness of the gas in the ISM comes from molecular clouds and we 
allow for correlation between the SNR and the clouds in the analysis. \\
(iv) Results of the type shown in Figure 2 show a dependence of $\sigma$ on $\Delta l$ 
in the range $\Delta l: 1^\circ - 8^\circ$ which is very similar for all the models and
 the observed values.

Although the implication of the above is that the effect of the fluctuations in $N(H)$
(~both atomic and molecular~) is not likely to be dramatic, it is certainly finite.
Its magnitude will be greatest in the Inner Galaxy at low latitudes, $|b| < 2.5^\circ$,
where contributions from the 'far' Inner Galaxy are important.

Estimate for both the Inner and Outer regions have been made. For $N(HI)$ the estimated
 value of the 
logarithmic standard deviation is 0.027 using the well known column densities of $HI$
\cite{hunt}. For $N(H_2)$, data \cite{dame} give a value 0.026 referred to the 
whole column density (~$N(HI)+N(H_2)$~), leading to an overall value of 
$\sigma_{gas}$ = 0.037.
Comparison can be made with the observed value of $\sigma_{obs}$ for this $l,b$ region:
 0.080. Subtraction in quadrature leads to $\sigma$  = 0.071, i.e. a small, but finite 
reduction.

Support for our analysis comes from a study of the EGRET observations and analysis 
published prior to the latest observations used by us \cite{hunt,cill}. In this work, 
predictions were made for a model in which the column densities of gas were used 
together with an inferred CR intensity distribution which was correlated with the 
volume density of gas (~derived from the column density and assumed rotation curve~) 
but 'smoothed' spatially at the {\em kpc} level. The majority of the excursions in 
predicted intensity were thus due to the gas column density variations alone. The 
fluctuations of the observed intensities 
(~$E_\gamma > 1 GeV, |b| < 2^\circ, |l|: \pm90^\circ$~) about the prediction have 
$\sigma = 0.06 \pm 0.015$. This value is quite consistent with ours.

Turning to regions away from the Galactic Plane in the Inner Galaxy 
(~$|b| = 2.5^\circ-5^\circ$~) the corresponding value for $HI$ and $H_2$ is 
$\sigma_{gas} = 0.04$. In the Outer Galaxy the values are $\sigma = 0.025$ for 
$|b| < 2.5^\circ$ and 0.035 for $|b| = 2.5^\circ/5^\circ$. It will be noticed that 
in all cases the corrections will be small.  

Our contention is that the major cause of the fluctuations in gamma-ray intensity on 
the $4^\circ$ scale is the stochastic nature of the SNR in space and time and, 
probably, in energy output. The column density of gas undoubtedly fluctuates with 
{\em l} but the observed gamma-ray intensity fluctuations are much greater because of 
the CR intensity variations along the line of sight.
\begin{figure}[htb!]
\begin{center}
\includegraphics[height=15cm,width=15cm]{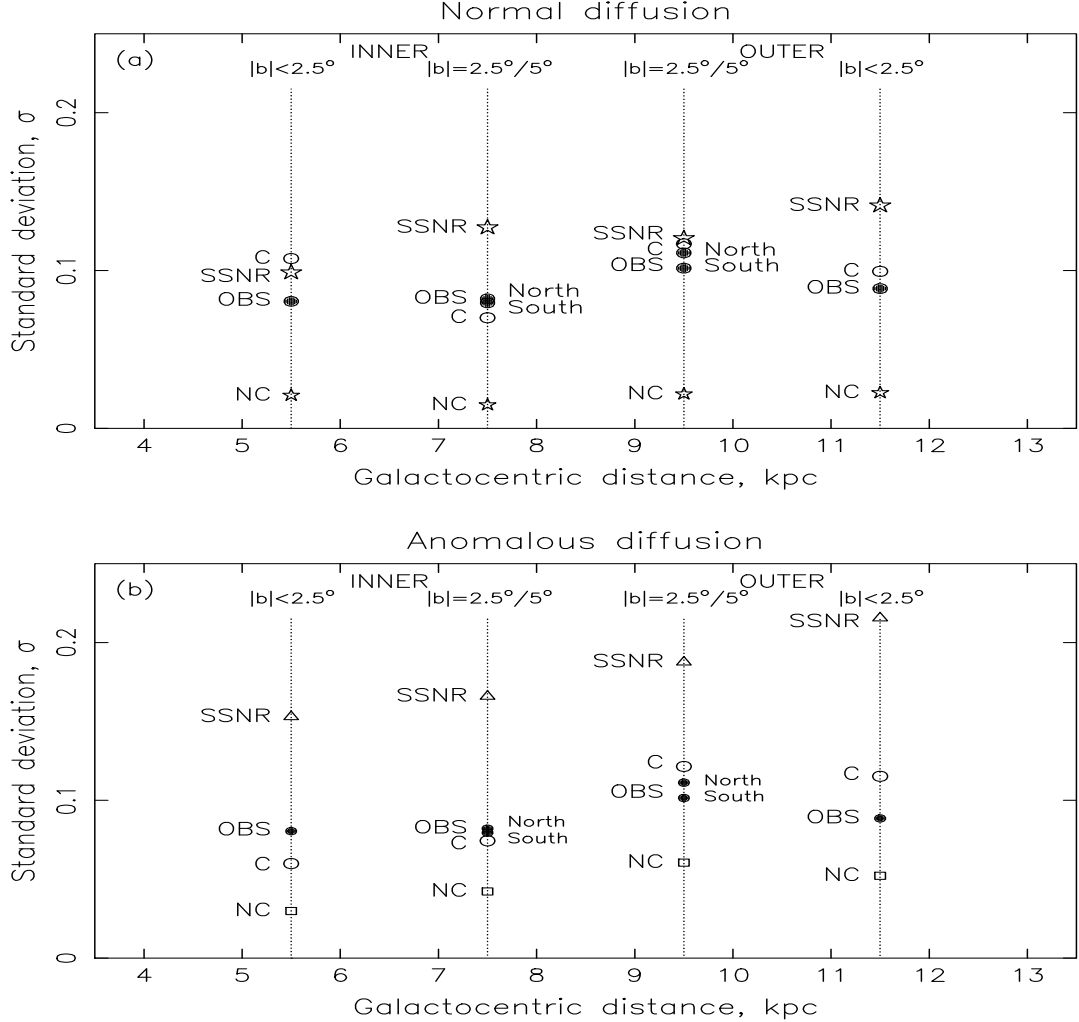}
\caption{\footnotesize Summary of standard deviations ($\sigma$-values) of 
gamma ray intensities about the best-fit: (a) - normal diffusion, (b) - anomalous 
diffusion. Small downward corrections have been 
applied to the OBS (~observed~)-values to allow for experimental noise. 
Model simulations are: NC - non-clustering, C - clustering, SSNR - with a frequency 
distribution of energies. Mean standard deviations are obtained by averaging 4-8 
independent samples. Errors are not indicated for clarity of the figures, but the 
distribution of $\sigma$ for different samples is broad due to the stochastic nature of
 SN explosions. Typical errors for anomalous diffusion are 0.002-(NC), 0.0015-(C) and 
 0.014-(SSNR) for the Inner Galaxy, 0.004-(NC), 0.004-(C) and 0.02-(SSNR) for the Outer
 Galaxy. For normal diffusion the errors are by 2-5 times less. The Galactocentric 
distances are the means from which much of the gamma ray intensities are derived.}
\end{center}
\label{fig:grsky3}
\end{figure}    
\subsection{The model predictions}
Our 'usual' model \cite{EW1,EW3} comprises CR production by SNR, as described
in paragraph 1, and propagation with `anomalous diffusion'.  The
SNR were assumed to be all identical and independent in time and space (~i.e. NSNR~), 
within the constraints of the dependence of the mean SNR density on Galactocentric 
radius.  We denote this as NC (~non-clustering~), and the results for this situation 
are shown in Figure 3. This is the NSNR model, but with no clustering.

The SSNR model has the same {\em mean} energy content per SNR as in the previous case 
but a distribution of energies about the mean, with a frequence distribution of 
{\em log energy} having a {\em log standard deviation} 
of 1.2, following \cite{svesh}. The results for this type of super-supernovae are now 
indicated as SSNR in Figure 3.
 
Turning to the possibility of the clustering of NSN, our clustering model has 
the following idealised form.  At any one place there are SN in sets of 10, 
distributed at random, in time, over $10^{6}y$.  The results for this model `C'
 (~clustered~), are also shown in Figure 3. Here, the SN are identical. 

Finally, calculations have been made for the same situation, as indicated, for 
'normal diffusion' (ND) rather than \underline{our} standard: 'anomalous 
diffusion'. 
\subsection{Comparison of observations with the models}
\subsubsection{General remarks}
It is apparent that the observed $\sigma$ values are higher than expected for 
the NSN model without clustering - i.e. non-clustering (NC). For normal diffusion 
NC(ND) the discrepancy is bigger still. Below we examine various factors which 
influence our model predictions.
\subsubsection{Mixed primary mass composition}
The model predictions shown in Figure 3 relate to primary protons. However, the 
observed gamma rays are produced by all CR nuclei, i.e. by the mixed primary mass
composition. Primary nuclei are more efficient in producing gamma rays than protons 
of the same energy per particle due to the higher interaction cross-section and higher
number of interacting nucleons in the collision. At the same time fluctuations of the
gamma-ray intensity created by CR nuclei are lower than those for protons, therefore 
our model predictions should be reduced for the mixed primary mass composition. 
We calculated the fluctuations for the mixed composition with relative abundances
equal to 0.416 for protons, 0.265 for helium, 0.170 for CNO group, 0.149 for iron group
of nuclei \cite{bier}. The result is that we have to multiply our predicted
$\sigma$-values by 0.93, 0.86, 0.85, 0.88 respectively for the increasing 
galactocentric distances indicated in Figure 3. The uncertainty of this correction 
is within the range of 0.005 - 0.019. As a result of the correction the discrepancy 
between observed and expected $\sigma$-values for the non-clustering case (NC) with a 
mixed primary mass composition is bigger still.   

\subsubsection{Mode of diffusion}
It is relevant to examine the question of 'normal or anomalous 
diffusion ?'. The mode of diffusion has obvious implications for the present work.
Calculations have been made for all cases, i.e. non-clustering, clustering and SSNR
models using normal diffusion, instead of anomalous. The result is that the 
fluctuations and $\sigma$-values are lower for normal diffusion in all cases (~see 
Figure 3a compared with Figure 3b~), thus the difference between the non-clustering 
case and observations will be even larger. 

\subsubsection{Super-supernovae}
Examination of Figure 3 shows that the $\sigma$-values for SSNR with a distribution of
 explosion energies is higher than for clustering (C) and for the observed fluctuations
 (OBS). It is due to the basic assumption about the fluctuations of the explosion 
energy. Added to the fluctuations caused by the stochastic distribution of SN in space 
and time these fluctuations result in a rise of the total fluctuations. 
Inevitably, it is possible to 'dilute' the SSNR values by assuming that only a fraction
of the SNR are of such high energy. According to \cite{svesh} the fraction of such 
SSNR is equal to 0.2 for SNIbc and 0.1 for SNIIn, but we regard it as a 
quantity 'to be determined'. 

\subsubsection{Clustering of supernovae and the presence of SSNR}
Although at first sight the case for some clustering of SNR at all
distances is strong there is a complication that must be allowed
for.  This concerns the correlation of SNR with target gas,
particularly in molecular form (~principally $H_{2}$~).  This
correlation is on linear scales smaller than the kpc-scale already
allowed for.  Our $4^{\circ}$ bin of longitude would correspond to
a linear scale of 280~pc at a distance of 4 kpc (~the typical
distance to the production region in the Inner Galaxy for
$|b|<2.5^{\circ}$~).  Such a distance would embrace likely
distances of molecular clouds (MC) from some, at least, of the SNR.

The magnitude of this correlation is not clear but some progress
can be made.  The list of known SNR \cite{green} has only 3\% of the
entries mentioning adjacent MC but this fraction will increase
with increasing SNR-MC distance.  Inspection of the Galactic maps
of MC \cite{dame} in the Inner Galaxy shows a mean separation of about
500pc; thus a correlation factor of about 0.5 would appear to be
indicated.  For the general case, it appears that the relation
between the various quantities for the situation where there is
partial clustering of the SN (~coefficient $f$~) and a partial correlation
between SNR and molecular gas (~coefficient 0.5~) is:
\begin{eqnarray}
\sigma_{obs}^2-\sigma_{gas}^2=(1-g)\{(1-\delta)[f(R_c\sigma_c)^2+(1-f)(R_{nc}\sigma_{nc}^2]+\delta(R_{ssnr}\sigma_{ssnr})^2\} \nonumber \\
+g[(1-\delta)\frac{R_{c}\sigma_c^2+R_{nc}\sigma_{nc}^2}{2}+\delta(R_{ssnr}\sigma_{ssnr})^2]
\end{eqnarray}
where $\sigma_{obs}$ is the total, observed (corrected) standard deviation, 
$\sigma_{gas}$ characterizes the fluctuations of the column density of the gas (~as
deived in \S3.2~), 
$\sigma_{nc}$, $\sigma_c$ and $\sigma_{ssnr}$ are the expectations for non-clustering, 
clustering and SSNR models respectively and the mixed primary mass composition, 
$R_{nc}$, $R_c$ and $R_{ssnr}$ - ratios of
the model to the observed gamma-ray intensities for the same three models, $g$ is the 
fraction of gas in molecular form and $\delta$ is the fraction of SSNR among the total 
SNR. The first term on the {\em rhs} of equation (1) relates to SNR in the ordinary 
interstellar medium, the second term - to SNR inside MC.
The equation satisfies the lower limit 
$\sigma_{obs}=\sigma_{nc}~\textrm{when}~f=\delta=g=0$

In the calculations it is assumed that SSNR are so rare that they are not clustered, 
  whereas half of the ordinary SNR (NC) are clustered in MC. 
The values of $g$ are, for the approximate median D-values relating to Figure 3: 0.5, 
0.3, 0.1 and 0.1. In fact, different models give different absolute gamma-ray 
intensities, but they can be easily normalised to the experimental data multiplying 
the explosion energy by a constant. The value of relative fluctuations $\sigma$ in 
this operation remains unchanged. Therefore, we adopted all $R_{nc}=R_c=R_{ssnr}=1$.

The fraction $f$ is therefore the only unknown variable and it can be calculated from 
equation (1). The results are shown in Figure 4.
\begin{figure}[htb!]
\begin{center}
\includegraphics[height=15cm,width=15cm]{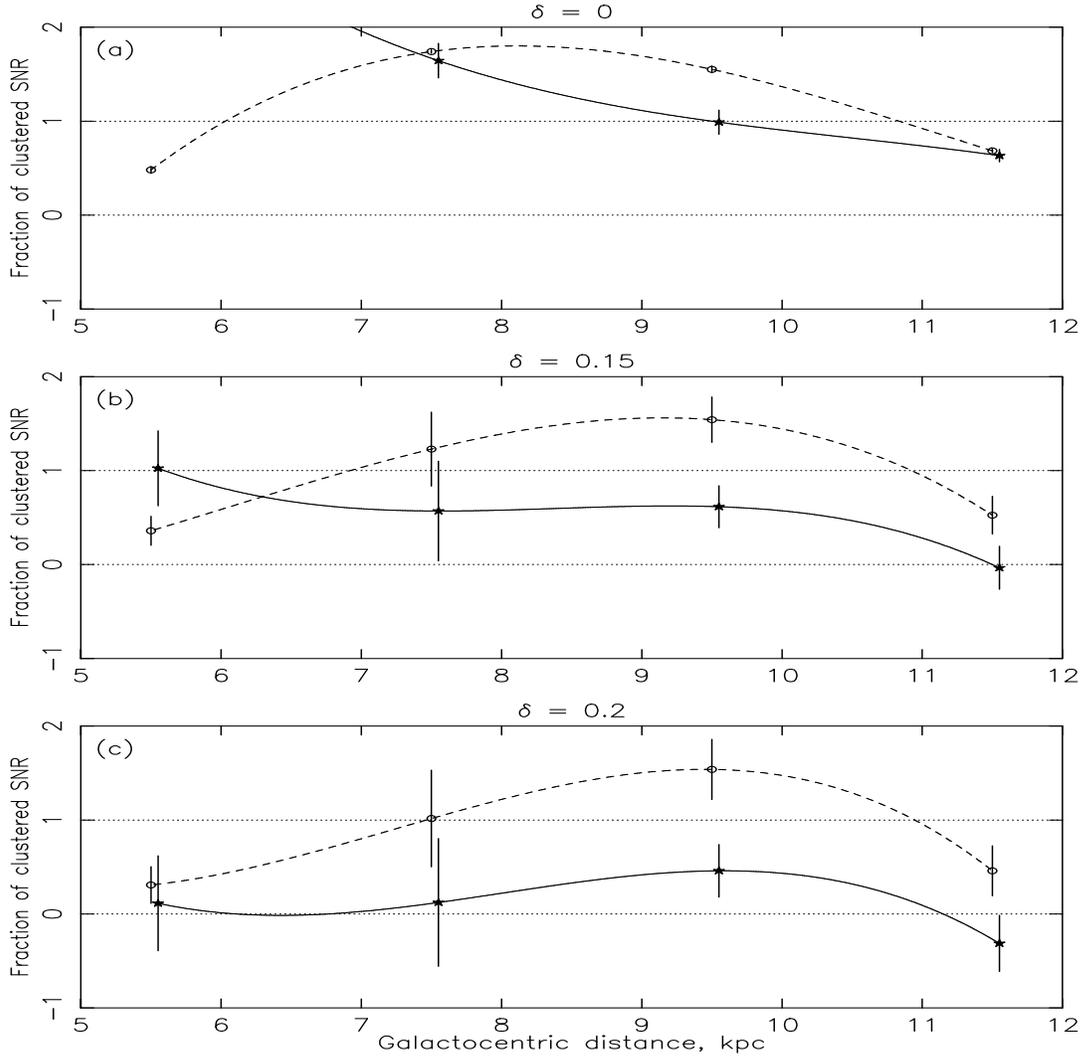}
\caption{\footnotesize The fraction of the clustered SNR as a function of 
galactocentric distance for normal (~$\circ$~) and anomalous (~$\star$~) diffusion and
for 3 SSNR fractions: (a) - $\delta=0$, (b) - $\delta=0.15$, (c) - $\sigma = 0.2$.
The lines connecting the points are 3-degree polynomial fits to the data. The abscissae
 for the two curves are slightly displaced with respect to each other to avoid an 
overlap of the errors. For $\delta = 0$, the first point (~extreme left~) for anomalous
 diffusion is at $f = 3.21$.}
\end{center}
\label{fig:grsky4}
\end{figure}    

\section{Discussion}
The situations for the two models can be considered in turn.

For the NSNR model, the model used by us in much of our work, considerable clustering 
is needed in the Inner Galaxy. Indeed, with $f > 1$ the clustering must exceed that 
adopted by us, viz. 10 SN per $10^6y$ coincident in position. Although not impossible, 
it is unlikely that the necessary higher number (~20-30 ?~) is present, even in the 
congested region of the Giant Molecular Ring at $D \approx 4kpc$.

Instead, we incline towards the SSNR model. Inspection of Figures 4b and 4c shows that 
in the important Inner Galaxy region, with $\delta = 0.2$, the bulk of the work is done
by SSNR and little by SNR. A value of $\delta$ in the range 0.15 to 0.2 is indicated, 
 with 0.15 being more physical. In the Outer Galaxy , the likely range is 0.0 to 0.15.

Certainly, the values of $f$ shown in Figure 4 have big systematic and statistical 
errors due mainly to the simplified model of clustering and big fluctuations caused by 
the stochastic nature of SN explosions. The estimates of these errors are indicated in 
Figure 4.     

\section{Conclusions}
The fluctuations in the intensity of gamma rays above 1 GeV in and near the Galactic 
Plane give information about the mode of production of cosmic rays in their sources, 
and their mode of propagation. 

We have adopted alternative models of both sources ( supernovae of a variety of 
energies and SN of unique energy but with varying degrees of 'clustering' ) and 
diffusion ( our 'standard' anomalous
diffusion, and normal ). For 'normal diffusion', although at $D \sim$ 5.5 and 11.5 kpc 
it is possible to achieve 'reasonable' results, for $7 < D < 11$ kpc (~the region 
nearest  the sun~) this is not possible. Whatever the value of $\delta$, the fractions 
are unreasonably high. Perhaps this argument can be used to point against 'normal' 
diffusion being applicable for CR of the energies in question ?

Bearing in mind our general arguments that clustering is more common in the Inner 
Galaxy it would appear that $\delta \sim$ 0.15 is favoured (~Figure 4b~). We would then
 have a constant fraction of SN of the SSN variety, with none of them clustered, but 
the normal SN clustered only in the solar vicinity and the Inner Galaxy. 

It seems that the solution put forward is astrophysically reasonable, for the following
 reasons. Concerning standard SNR, these come from stars of modest mass produced in 
Molecular Clouds. These clouds are composed of individual clumps in which star 
production (~and subsequent SN~) occurs. In the Inner Galaxy, where the overall MC 
masses are bigger \cite{kutn} there will be more clusters than in the Outer. For SSN,
the fractional number per clump of the required massive progenitor stars is probably 
the same in the Inner and Outer Galaxy. In view of continuing problems with theories of
 star formation [~T.J.Millar, private communication~] even this conclusion cannot be 
regarded as completely firm.  

{\bf Acknowledgments}

The Royal Society and the University of Durham are thanked for financial support.  

\newpage
\begin{center}
\bf{Table 1}
\end{center}
\begin{center}
\begin{tabular}{|c|c|c|c|c|} \hline
&$\triangle\ell$&No. of entries&Mean&$\sigma$\\ \hline
Inner&$1^{\circ}$&161&0.00295&0.087\\
&$2^{\circ}$&80&0.00225&0.086\\
&$4^{\circ}$&40&0.00175&0.081\\
&$8^{\circ}$&20&0.000500&0.074\\
Outer&$1^{\circ}$&180&-0.00272&0.098\\
&$2^{\circ}$&90&-0.00156&0.096\\
&$4^{\circ}$&45&-0.00167&0.091\\
&$8^{\circ}$&23&+0.00022&0.081\\\hline
\end{tabular}
\end{center}
\footnotesize{Table 1. Results for the distribution of the differences between the 
EGRET intensities and the 9- degree polynomial fit: nos. of entries,
means and RMS ($\sigma$-) values.  All values relate to the
logarithm of the intensity.  Latitude range,
$|b|<2.5^{\circ},~E\gamma>1$ GeV.  The histograms are given in Figure 2.}

\end{document}